\documentclass{elsart}

\usepackage{graphics}
\usepackage{epsfig}
\usepackage{subfigure}
\usepackage{amsmath,amsfonts,amssymb,graphicx,times}

\begin{document}

\begin{frontmatter}

\title{Information Filtering via Implicit Trust-based
Network}

\author[1]{Zhuo-Guo Xuan},
\author[1]{Zhan Li},
\author[1,2]{and Jian-Guo Liu}\ead{liujg004@ustc.edu.cn}

\address[1]{Institute of Systems Engineering, Dalian University of
Technology, Dalian 116024, People's Republic of China}
\address[2]{Research Center of Complex Systems Science, University of
Shanghai for Science and Technology, Shanghai 200093, PR China}

\begin{abstract}
Based on the user-item bipartite network, collaborative filtering
(CF) recommender systems predict users' interests according to their
history collections, which is a promising way to solve the
information exploration problem. However, CF algorithm encounters
cold start and sparsity problems. The trust-based CF algorithm is
implemented by collecting the users' trust statements, which is
time-consuming and must use users' private friendship information.
In this paper, we present a novel measurement to calculate users'
implicit trust-based correlation by taking into account their
average ratings, rating ranges, and the number of common rated
items. By applying the similar idea to the items, a item-based CF
algorithm is constructed. The simulation results on three benchmark
data sets show that the performances of both user-based and
item-based algorithms could be enhanced greatly. Finally, a hybrid
algorithm is constructed by integrating the user-based and
item-based algorithms, the simulation results indicate that hybrid
algorithm outperforms the state-of-the-art methods. Specifically, it
can not only provide more accurate recommendations, but also
alleviate the cold start problem.
\end{abstract}

\begin{keyword}
Recommender systems \sep Bipartite networks \sep Collaborative
filtering. \PACS 89.75.Hc\sep 87.23.Ge\sep 05.70.Ln
\end{keyword}

\end{frontmatter}

\section{Introduction}
Information exploration is one of the results of internet and social
network development. The swift and violent growth of information on
the Internet makes it more and more difficult for users to find
available and useful portions \cite{1}. How to help the users find
out the relevant information or products by using the user-item
bipartite network is a promising way to solve the information
overload problem \cite{2,PNAS,NJP2009}. Search engineering and
recommender systems are two effective tools to help users filter out
what pieces are relevant to their tastes. However, search
engineering presents exactly same list to the same keywords
regardless of users' interests, habits and the history behavior
information. Recommender systems filter out the irrelevant
information and recommend the potentially interesting items to the
target users by analyzing their interests and habits through their
history behaviors, which have been successfully applied in a lot of
e-commercial web sites \cite{3,lu}.

Collaborative filtering (CF) algorithm is one of the most successful
technologies for recommender systems, which firstly identifies the
target user's neighbors whose interests or habits are similar and
then presents the recommendation list according to the neighbor
users' history selections \cite{2,Liu2009,Liu2010}. Recently, the
similar idea has been applied to the items. Generally speaking, CF
algorithms can be systematically classified as user-based and
item-based \cite{1}. User-based methods, regarding each user's
ratings as a vector, measure the similarity between the target user
and those like-minded people and predict the target user's rating
for the target item according to the history preferences. User-based
CF algorithms have been investigated extensively \cite{Dun2009}. For
example, Herlocker {\it et al.} \cite{2} proposed an algorithmic
framework referring to user similarity. Luo {\it et al.} \cite{5}
introduced the {\it local user similarity} and {\it global user
similarity} concepts based on surprisal-based vector similarity and
the concept of maximum distance in graph theory. When the number of
items is approximately constant, it is better to give the prediction
according to items' similarity network. Item-based methods,
regarding each item's ratings as a vector, measure the similarity
between the target item and other items and predict the target
rating relying on users' preferences in history. Because of less
updates for average items and comparatively static state, the
item-based approaches are superior. Sarwar {\it et al.} \cite{6}
proposed item-based CF algorithm by comparing different items.
Deshpande {\it et al.} \cite{7} proposed item-based top-$N$ CF
algorithm, in which items are ranked according to the frequency of
appearing in the set of similar items and the top-$N$ ranked items
are returned. Recently, Gao {\it et al.} \cite{8} incorporated the
user ranking information into the computation of item similarity to
improve the performance of item-based CF algorithm.

In the previous work, a lot of rating information wasn't taken into
consideration to compute the user or item similarity, such as
average ratings, rating ranges, the number of users' common rated
items and so on. We argue that, however, these information should be
taken into account to measure users' relationship.

When some new users enter into a recommender system, they only give
ratings to a few items. Analogously, when some new items are added
in the system, they only receive ratings from a few users, which is
named {\it cold start problem}. It's very hard to give high quality
prediction based on less of history selection information. In order
to solve the cold start problem, some researchers attempt to
integrate user-based and item-based CF methods to avoid the
limitation of one single algorithm. For instance, Kim {\it et al.}
\cite{9} built united collaborative error-reflected models that
reflect the average pre-prediction errors of user neighbors and of
item neighbors. Jeong {\it et al.} \cite{10} proposed an iterative
semi-explicit rating method that extrapolates unrated elements from
similar users and items in a semi-supervised manner. Besides, Lee
{\it et al.} \cite{11} used ratings data horizontally and vertically
to make two-way cooperative prediction for CF algorithm and thus
categorized four possible cases of predictions, namely equivalent
case, user-winning case, item-winning case and prediction-impossible
case. Empirical experiments show integrating user-based and
item-based methods could enhance the performance greatly.

%The trust-based recommender system are constructed according to the
%investigated report that we prefer to accept the recommendations
%from our trust friends but not the ideally related items according
%to their history preferences.

Recently, trust-based mechanism is introduced to alleviate the
cold-start problem. Some of e-commerce web sites, such as Epinions,
eBay and etc., try to apply trust mechanism to recommend products to
consumers. In these web sites, the trust mechanism is implemented by
collecting explicit or implicit trust statements. Explicit trust
statements need users to indicate the trust values to their friends
\cite{12}. Massa {\it et al.} \cite{4} suggested the explicit
trust-aware CF recommender systems by searching trust neighbors in
depth-first way according to trust propagation. Jamali {\it et al.}
\cite{12} built a model, named TrustWalker, by random walk in social
trust network to find trust neighbors who have rated the target item
or similar items. However, the above trust-based recommendation
algorithms need explicit trust statements expressed by users, which
are time-consuming and probably expose users' privacy. Therefore,
some implicit trust methodologies are proposed
\cite{13,14,15,16,17}. O'Donovan {\it et al.} \cite{13} proposed
computational models by implicit trust based on initial ratings,
which only studied the effects of the errors between predicted
ratings and actual ratings. Moreover, Kwon {\it et al.} \cite{14}
created a multidimensional credibility model for neighbor selection
in CF algorithm by deriving source credibility attributes (i.e.,
expertise, trustworthiness, similarity and attraction) and
extracting each consumer's importance weight. Li {\it et al.}
\cite{15} applied fuzzy logic and inference to support peer
recommendation service. Jeong {\it et al.} \cite{16} developed user
credit-based CF methods which incorporate the information of each
user's credit on rating items to compute the aggregation weight.
What's more, Lathia {\it et al.} \cite{17} proposed the trusted
$k$-nearest recommenders algorithm which allows users to learn who
and how much to trust others by evaluating the utility of the rating
information they have received.

In previous work, the users' rating habits wasn't taken into
account, such as average ratings, rating ranges, the number of
common rated items and so on. We argue that these factors are very
important and could be used to measure the implicit trust-based
similarities between users or items. In this paper, by constructing
the implicit trust-based network, we present three algorithms, say
user-based, item-based and hybrid algorithms. The simulation results
indicate that these factors are important and the hybrid algorithm
outperforms the state-of-the-art methods and performs very well to
the cold-start problem.

The following sections are organized as follows: Section 2, we
describe the definition and measurement how to calculate the
implicit trust-based user or item similarity, and the corresponding
algorithms are also introduced. In Section 3, the simulation
experiments on MovieLens, Netflix and Jester data sets are
investigated and the results are analyzed in detail. Finally, the
conclusions are presented and future work is discussed in Section 4.

% Results and Discussion can be combined.

\section{Collaborative filtering algorithms based on implicit trust-based network}

\subsection{User-based Collaborative Filtering Algorithm}

\subsubsection{Definition of implicit trust-based user correlation network}
The meaning of implicit trust-based users can be found in some
previous work. For instance, O'Donovan {\it et al.} \cite{13}
supposed that the trustable partners have similar tastes and
preferences to the target user and they should be trustworthy in the
sense that they have a history of making reliable recommendations,
whereas Kwon {\it et al.} \cite{14} conceived that trustable
neighbors have high expertise, trustworthiness, similarity, etc. In
addition, Jeong {\it et al.} \cite{16} set the trust-based user as
the similarity of voting a rating score with others. Hereinafter,
{\it Trust} in recommender system is defined in the following way.
When a user agrees with another user about quality of certain
products, she probably builds trust relationship with another, which
further means that their similar opinions might be inferred in some
ways.

In this paper, a trust-based user is defined as the user who has the
implicit trust relationship with the target user. Since trust in
e-commerce largely depends on similar views between users, implicit
trust in this paper can be explained as the similarity of their
opinions and interests on products, which are involved in average
ratings, rating ranges and the number of common rated items.

\subsubsection{Implicit Trust Measurement}
In recommender systems, users express their opinions in the form of
reviews, ratings, etc. Therefore, we could analyze their interests
from different angles to build correspondingly implicit trust among
them. In this paper, three factors are taken into consideration to
learn about their interests: 1) users' average ratings, 2) the
ranges of their ratings, 3) their common experience. The details are
discussed as follows:

1) {\bf Average ratings.}

Every user has his/her independent rating schema, i.e. his/her
average rating in a recommender system, as a result of his/her
distinct personal characteristics. When users pay attention to their
favorable items, they may express their opinions by variant ratings.
In consequence, their independent rating schemas are generated,
which reflect their own characteristics. In traditional CF
recommendation algorithms, the rating schema is presented as the
average rating of a user. For example, lots of measurements are
proposed to define the similarities among users, such as Pearson
correlation coefficient \cite{2}, adjusted cosine similarity
\cite{6}, mass diffusion \cite{Zhou2007,Liu2009}, heat conduction
\cite{PNAS,Liu2011} and so on. Empirical studies show that these
measurements with average rating get better results than those
without average rating (e.g., cosine similarity) \cite{18}. In
mathematics and statistic domains, average rating reflects the
general level and the central tendency. Accordingly, in recommender
systems, those measurements are used to analyze how far users'
ratings are away from their average ratings and how their ratings
evolve. In other words, whatever the ratings are, if only the
differences and extents between users come close, the users are
considered similar. In this article, average ratings are taken into
account to measure the implicit trust values between users.

2) {\bf Rating ranges}.

The range of ratings given by a user is probably different for
another due to the diversity of users' habits, mood and contexts. In
the practical evaluation, some pessimistic users under bad mood and
contexts fall into the habit of giving low ratings for all items. On
the contrary, some positive users under good mood and contexts are
accustomed to giving high ratings for all items. Since the users do
not belong to the standard-rating sets, they should be treated
specifically. Therefore, the range of ratings for every user should
be taken into consideration when implicit trust weight is
calculated.

3) {\bf Common rated items}

We suppose the more information we receive from one person, the more
we know about her. Analogously, in recommender systems, users'
experience is supposed to be stressed. In recommender systems, the
common experience of users that they contribute to recommendation
should be observed in order to improve the performance of
recommender systems. For example, for the target user $u$ and two
neighbors, say $v$ and $w$, suppose the similarities between $u$ and
$w$, $v$ are equal, but user $u$ has more common rated items with
$v$ than $w$, therefore, it is reasonable to believe the similarity
between $u$ and $v$ is stronger. In our algorithm, common experience
between the target user and trustable neighbors is employed
entirely.

\begin{figure}[ht]
\center\scalebox{0.5}[0.5]{\includegraphics{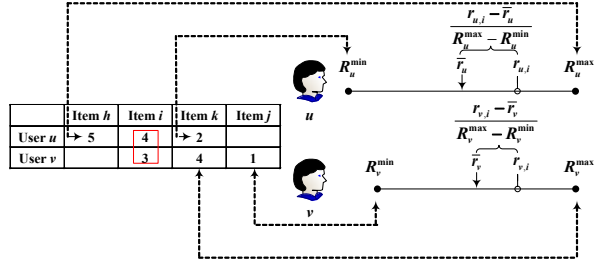}}
\caption{(Color Online) Implicit trust-based user correlation. For
user $u$ and $v$, their common experience is co-rated item $i$ and
$k$. For item $i$ (red rectangle), user $u$ gives normalized
distance from concrete rating value to average rating within his
rating range, and so does user $v$. The definition could be used to
measure the complement of absolute difference between the two users'
distances combining their common experience.} \label{Fig1}
\end{figure}

The main principle of implicit trust-based user correlation related
to the mentioned three factors is shown in Fig.\ref{Fig1}. For user
$u$ and $v$, their implicit trust-based correlation is calculated
based on their average ratings, say $\overline{r}_u$ and
$\overline{r}_v$, rating ranges, say $R_u^{\max}-R_u^{\min}$ and
$R_v^{\max}-R_v^{\min}$ and the number of their common rated items
$n$.

Considering the above three factors, we present the formulation to
calculate implicit trust between user $u$ and $v$:
\begin{equation}
C^U(u,v) =
\frac{1}{1+e^{-\frac{n}{2}}}\Big(1-\frac{1}{2n}\sum_{i=1}^n|\phi(u)-\phi(v)|\Big)
\end{equation}
where $\phi(u) =
\frac{r_{ui}-\overline{r}_u}{R_u^{\max}-R_u^{\min}}$ and $n$ is the
number of common rated items for user $u$ and $v$. The sigmoid
function, $1/(1+e^{-\frac{n}{2}})$, is used to rectify weight by the
number of common rated items, $n$, which has ever been distinctly
used to adjust Pearson Correlation coefficient \cite{12}.

\subsubsection{Prediction Based on Implicit Trust-based User Correlation Network}
In this paper, $K$-nearest neighbors of the target user are
evaluated to investigate the effect of implicit trust-based
correlation on cold start problem. Afterwards, the predicted rating,
$\widehat{r}_{uj}^U$ from user $u$ to the target item $j$ is given
according to the following formulation.
\begin{equation}
\widehat{r}_{uj}^U=\overline{r}_u+\frac{\sum_{v\in
\Gamma_u}C^U(u,v)(r_{vj}-\overline{r}_v)}{\sum_{v\in
\Gamma_u}C^U(u,v)}
\end{equation}
where $\Gamma_u$ is a set of the nearest neighbors of user $u$, and
$C^U(u,v)$ is the implicit trust-based correlation between user $u$
and $v$ obtained by Eq.(1).

\subsection{Item-based Collaborative Filtering Algorithm}
Introducing a similar idea on the item correlation definition, the
effect of the implicit trust-based correlation on item-based CF
algorithm is investigated.

\subsubsection{Definition of implicit trust-based items}
When we are satisfied with products that we have purchased, we
usually place them in trusted zone. Perhaps, in the future, we will
buy them again. On the contrary, if we complain about some bad
products, we place them in restricted zone and we may never buy them
again.

In this paper, items based on implicit trusts are considered relying
on proximity of items that a user has evaluated in her history. From
this point of view, trusted items can be explained as the items that
are close to those that one user trusts. In other words, while a
user set a certain item in her trusted zone, the trust-based items,
in terms of intrinsic attributes, accepted degrees, rating values
and common popularity, are very similar to it. The process to search
implicit trust-based items is to analyze all users' opinions about
these items.

\subsubsection{Implicit trust measurement}
In the paper, like implicit trust-based user correlation definition,
three factors are referred to compute items' implicit trust-based
relationship, which can be described as: 1) the internal or
intrinsic attributes of an item, 2) the accepted degree of an item,
3) the common rated times between any pairs of items. The detail is
described as follows:

1) {\bf Intrinsic attributes of an item}

The internal attributes of an item determine all users' opinions
about it. In other words, the average rating reflects intrinsic
attributes of the item. If the quality of an item is good, users
generally like it and give it high ratings, and vice versa. The more
users have evaluated an item, the closer the average rating is to
the internal characteristics of the item. The average rating implies
all users' opinions about the item.

We primarily pay attention to the distance from concrete rating to
average rating. That means, the nearer to average rating the
concrete rating value is, the more trustworthy an item is. In a
word, average rating plays a significant role to implement
recommendation based on implicit trust-based items.

2) {\bf Accepted degree of an item}

The accepted degree of an item can be observed from two
perspectives, minimum rating and maximum rating, which can be
inferred from the rated range of the item. To an item, the minimum
rating shows how bad the item a user thinks and the maximum one
shows how good the item she considers. In brief, the minimum and
maximum ratings describe the accepted degree of an item derived from
all users' opinions. For instance, if a movie is rated with low
ratings.

3) {\bf Common rated times between items}

The number of users who have commonly rated the items could affect
the trustworthy levels of the items.

The more users give high ratings to two items, the more correlated
these items are. Generally, the number of common rated times between
the target item and its implicit trust-based neighborhood items
should be taken into account.

\begin{figure}[ht]
\center\scalebox{0.5}[0.5]{\includegraphics{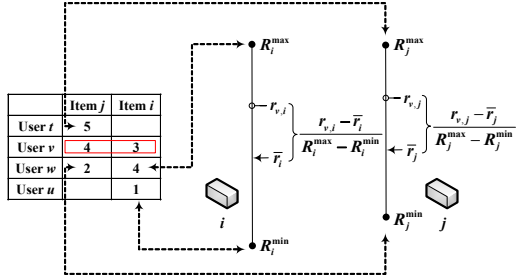}}
\caption{(Color Online) Three factors affecting implicit trust
weight of credible items. For item $i$ and $j$, their common
popularity is the aggregation of over $v$'s and $w$'s ratings. For
user $v$ (red rectangle), item $j$ gets normalized distance from
concrete rating value to average rating within its rating range, and
so does item $i$. The goal is to aggregate the complement of
absolute difference between the two items' distances combining their
common popularity.} \label{Fig2}
\end{figure}

The core principle of implicit trust-based item correlation is
depicted in Fig.\ref{Fig2}. For item $i$ and $j$, the intrinsic
attributes are denoted as their average ratings $\overline{r}_i$ and
$\overline{r}_j$ respectively. The differences between maximum and
minimum ratings are denoted as $R_i^{\max}-R_i^{\min}$ and
$R_j^{\max}-R_j^{\min}$. The number of common rated times is set as
$m$.

Therefore, the following formulation could be given,
\begin{equation}
C^I(i,j) =
\frac{1}{1+e^{-\frac{m}{2}}}\Big(1-\frac{1}{2m}\sum_{v=1}^m|\phi(i)-\phi(j)|\Big)
\end{equation}
where $\phi(i) =
\frac{r_{vi}-\overline{r}_i}{R_i^{\max}-R_i^{\min}}$ and $m$ denotes
the number of users who have rated both item $i$ and $j$. The
sigmoid function, $1/(1+e^{-\frac{m}{2}})$, is used to rectify
weight by common users.

\subsubsection{Prediction Based on Implicit Trust-based Item Correlation Network}
To investigate the effect of implicit trust-based item correlation
network on users' cold start problem, the $K$-nearest neighbors are
evaluated in this paper. The predicted rating from user $u$ to item
$j$ is given according to the following item-based CF algorithm.
\begin{equation}
\widehat{r}_{uj}^I=\overline{r}_j+\frac{\sum_{i\in
\Gamma_j}C^I(i,j)(r_{ui}-\overline{r}_i)}{\sum_{i\in
\Gamma_j}C^I(i,j)}
\end{equation}
where $\Gamma_j$ is a set of the nearest neighbors of item $j$, and
$C^I(i,j)$ denotes the implicit trust weight between item $i$ and
$j$ by Eq.(3).

\subsection{Hybrid algorithm}
Traditional CF algorithm encounters cold start problem because of
data set sparsity, which can be further divided into cold start
users and cold start items \cite{9}. A cold start user indicates the
new user who has participated in recommendation but has expressed
few opinions. In this situation, it is often the case that there is
no intersection at all between two users, and it is difficult to
calculate the user similarity based on common rated items. Even when
the computation of similarity is possible, it may not be very
reliable because of the insufficient information available. A cold
start item is caused by the new item. In the CF-based recommender
systems, this item cannot be recommended due to insufficient user
opinions. The simulation results indicate that the hybrid algorithm
could not only greatly enhance the accuracy, but also effectively
solve the cold start problem.

In this paper, to alleviate the cold start problem, we present a
hybrid recommendation algorithm by integrating implicit trust
user-based and item-based CF algorithms, where the predicted rating
is given in the following way
\begin{equation}
\widehat{r}_{uj}=(1-\alpha)\widehat{r}^U_{uj}+\alpha\widehat{r}^I_{uj},
\end{equation}
where $\widehat{r}^U_{uj}$ is the prediction rating based on
user-based CF algorithm in Eq.(2), $\widehat{r}^I_{uj}$ is the
prediction rating based on item-based CF algorithm in Eq.(4), and
$\alpha$ is a tunable parameter whose range is [0,1]. When $\alpha
=0$, the hybrid algorithm degenerates to the user-based algorithm,
and it becomes the item-based CF algorithm when $\alpha = 1$. We can
adjust value to control the ratios from the above two algorithms and
find the optimum solution.

\section{Simulation Results}

\subsection{Data Description and Statistical Properties}
In this paper, our simulation experimental data comes from MovieLens
\footnote{http://www.Movielens.com},
Netflix\footnote{http://www.netflix.com} and Jester. The Movielens
data is collected by the GroupLens Research Project during the
seven-month period from September 19th, 1997 through April 22nd,
1998. The dataset consists of 100,000 ratings from 943 users on
1,682 movies and rating scale is from 1 (awful) to 5 (must see),
which has been cleaned up so that users who had less than 20 ratings
or did not have complete demographic information were removed from
this dataset. The Netflix and Jester data are random samples of the
whole records of user activities in Netflix.com and Jester, in which
the Netflix data consists of 10000 users, 6000 movies and 824802
links, and the Jester data has 2350 users, 100 jokes and 169,655
connections. Table gives the statistical properties of the test data
sets.

\begin{table}
\begin{center}
\caption{Basic statistics of the test data sets.}
\end{center}
\begin{center}
\begin{tabular} {ccccc}
  \hline \hline
   Data Sets      &  Users   & Objects & Links & Sparsity \\ \hline
   MovieLens      &  6,040    & 3,592    & 750,000 &  $3.46\times 10^{-2}$\\
   Netflix        &  10,000  & 6,000   & 701,947    & $1.17\times 10^{-2}$ \\
   Jester         &  2,350   & 100     & 169,655 & 0.7219\\
   \hline \hline
    \end{tabular}
\end{center}
\end{table}

\subsection{Evaluation metrics}
In order to measure the performances of the present algorithms, the
mean absolute error (MAE) \cite{19}, the root mean square error
(RMSE) \cite{12} and the hit rate (HR) are used.

\subsubsection{Mean Absolute Error}
MAE is the mean absolute difference between an actual and a
predicted rating value, which is generally used for the statistical
accuracy measurements in various algorithms. The smaller MAE an
algorithm achieves, the better the experimental result is. The
metric MAE is defined as:
\begin{equation}
{\rm MAE}= \frac{\sum_{i=1}^{n_r}|\widehat{r}_i-r_i|}{n_r}
\end{equation}
where $\widehat{r}_i$ and $r_i$ represent the predicted and actual
rating respectively, and $n_r$ denotes the number of tested ratings.

\subsubsection{Root Mean Square Error}
RMSE has been typically used to measure the large errors in extreme
cases. Analogously, the smaller the value of RMSE an algorithm
obtains, the more precise the recommendation is. The metric RMSE is
usually defined as follows
\begin{equation}
{\rm RMSE}= \sqrt{\frac{\sum_{i=1}^{n_r}(\widehat{r}_i-r_i)^2}{n_r}}
\end{equation}

\subsubsection{Hit Rate}
The hit rate (HR) is also introduced to measure the accuracy of the
recommendation. Here, HR is defined as the ratio of the number of
hits (i.e., the fraction of the number of recommended items and
actually chosen items) to the size of the recommendation list. In
the information retrieval literature, it is usually equivalent to
the metrics {\it Precision} and {\it Recall}. The bigger the value
of HR is, the better an algorithm. Formally, HR is defined as
\begin{equation}
HR=\frac{H}{L},
\end{equation}
where $L$ is the length of recommendation list and $H$ is the
percentage of items in the test set existing in the top-$L$
positions of recommendation list.

\subsection{Experiment results analysis}
The implicit trust-based effects are implemented on user-based,
item-based and hybrid algorithms separately. Since the prediction
performance is influenced by the size of the $K$ nearest neighbors,
it is essential to determine a proper size of the nearest neighbors
Top $K$, where $K$ is set as 3, 5, 10, 15 and 20 respectively. Since
the typical length for recommendation list is ten items, our
experiments set $L$=10. The parameter $\alpha$ is adjusted in the
interval [0, 1] and the increment is 0.1.

\subsubsection{Performance of Implicit Trust-based Effect on User-based Algorithm}
In this section, we investigate the performance of the user-based CF
algorithm (denoted as IU-CF) and compare it against the performances
of classic user-based CF using well-known Pearson Correlation
coefficient (denoted as PCF) and adjusted cosine-based CF algorithm
\cite{6} (denoted as AC-CF).

\begin{figure}[ht]
\center\scalebox{0.6}[0.6]{\includegraphics{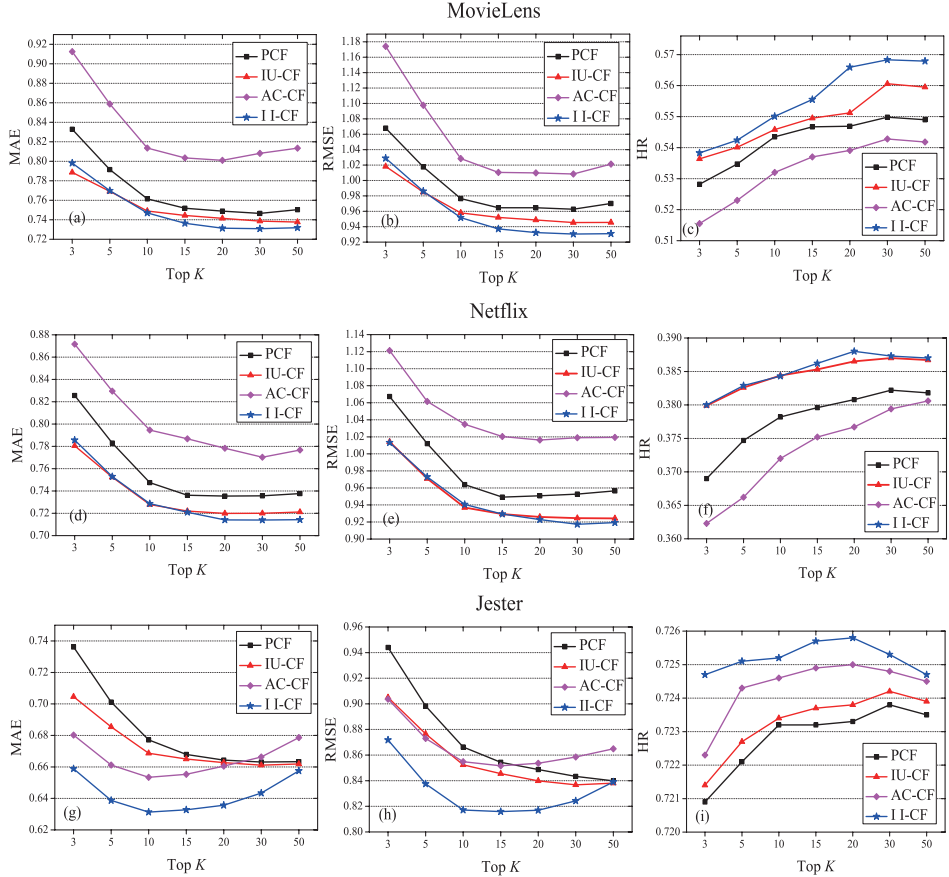}}
\caption{(Color Online) Comparison of results achieved by
Pearson-based (PCF), adjusted cosine-based (AC-CF), implicit
trust-based (IU-CF) and item-based CF (II-CF) algorithms. Note that
both IU- and II-CF have the smallest MAE, RMSE and highest HR for
Movielens, Netflix and Jester data sets.} \label{Fig6}
\end{figure}

Figure \ref{Fig6} illustrates the results of MAE, RMSE and HR for
PCF, AC-CF, IU-CF and II-CF algorithms respectively. The results
demonstrate that IU-CF and II-CF algorithms enhance the performance
of the initial two approaches, PCF and AC-CF. From Fig.~\ref{Fig6},
one can see that MAE of IU-CF algorithm has the lowest level in the
three algorithms. As the number of the nearest neighbors $K$
increases, the MAE curves of all four algorithms tend to decrease,
which implies that more neighbors can make better prediction
although computation and time complexity is high. The RMSE results
in Fig.\ref{Fig6} show that IU- and II-CF algorithms have the
smallest errors in the three algorithms while PCF algorithm gets
results with the largest errors. In other words, our approach can
predict more accurately than PCF and AC-CF algorithms. In addition,
the similar RMSE downtrend for all algorithms appears in
Fig.\ref{Fig6} as the growth of the sizes of user neighborhood.
Fig.\ref{Fig6} illustrates the results of HR of three algorithms. As
shown in Fig.\ref{Fig6}, at most neighborhood sizes, HR of IU- and
II-CF algorithms are remarkably better than the results of PCF and
AC-CF algorithms. Even though only a minority of neighbors
participate in prediction, the present IU- and II-CF algorithms
outperform the other two methods. And, when the number of nearest
neighbors increases, the curves of the three methods gradually
change upward and finally tend to become flat. From the results of
Fig.\ref{Fig6}, it can be concluded that the present user-based and
item-based approaches can provide better recommendations.

\begin{figure}[ht]
\center\scalebox{1.2}[1.2]{\includegraphics{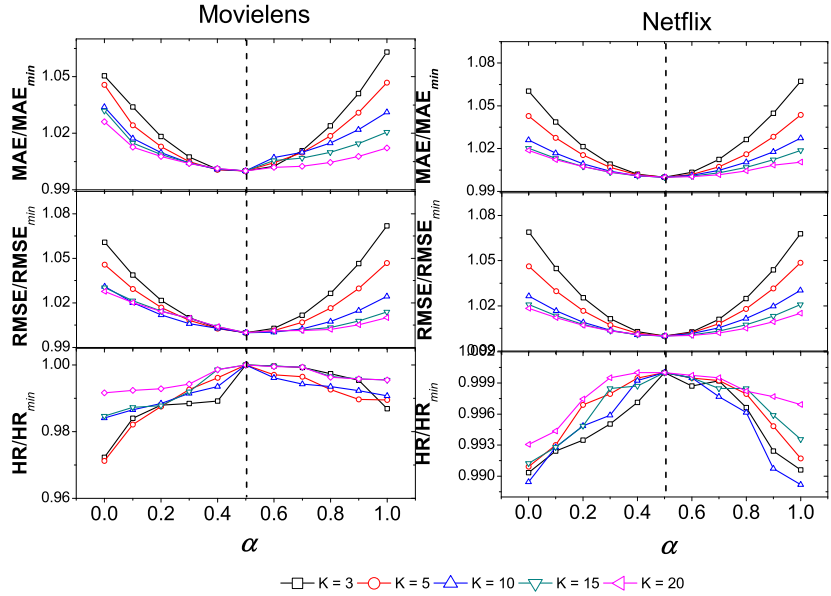}}
\caption{(Color Online) Comparison of setting different $\alpha$
values for HCF. Note that MAE and RMSE first fall down before
$\alpha$ =0.5 and then climb up after that for Movielens and Netflix
data sets. HR obeys reverse distribution with the boundary
$\alpha$=0.5. The conclusion is that the optimal $\alpha$ is 0.5 for
HCF.} \label{Fig72}
\end{figure}

\subsubsection{The performance of hybrid recommendation}
In this section, the effects of the implicit trust-based
correlations on hybrid recommendation (HCF) are investigated by
integrating the user-based and item-based CF algorithm. In the
experiment, we compare hybrid recommendation against the above two
pure algorithms with different values. Figure~\ref{Fig72} summarizes
the experiment results of MAE, RMSE and HR for HCF algorithm
according to the value $\alpha$ variation. We examine the HCF
results of the three metrics in order to choose optimal parameter
$\alpha$. In the experiment, the value is continuously changed in
the interval [0, 1] with the increment 0.1. From Fig.~\ref{Fig72},
MAE and RMSE apparently decrease as the value of increases from 0 to
0.5; after this point 0.5, the upward MAE and RMSE gradually appear
for Movielens and Netflix data sets. On the contrary, the metric HR
considerably ascends before the value 0.5 and after that it begins
to descend steadily. The optimal parameter for Jester data set is
not exactly 0.5, but also close to this value. The results indicate
that the optimal value is 0.5 no matter which metric is evaluated
for HCF.

\begin{figure}[ht]
\center\scalebox{0.6}[0.6]{\includegraphics{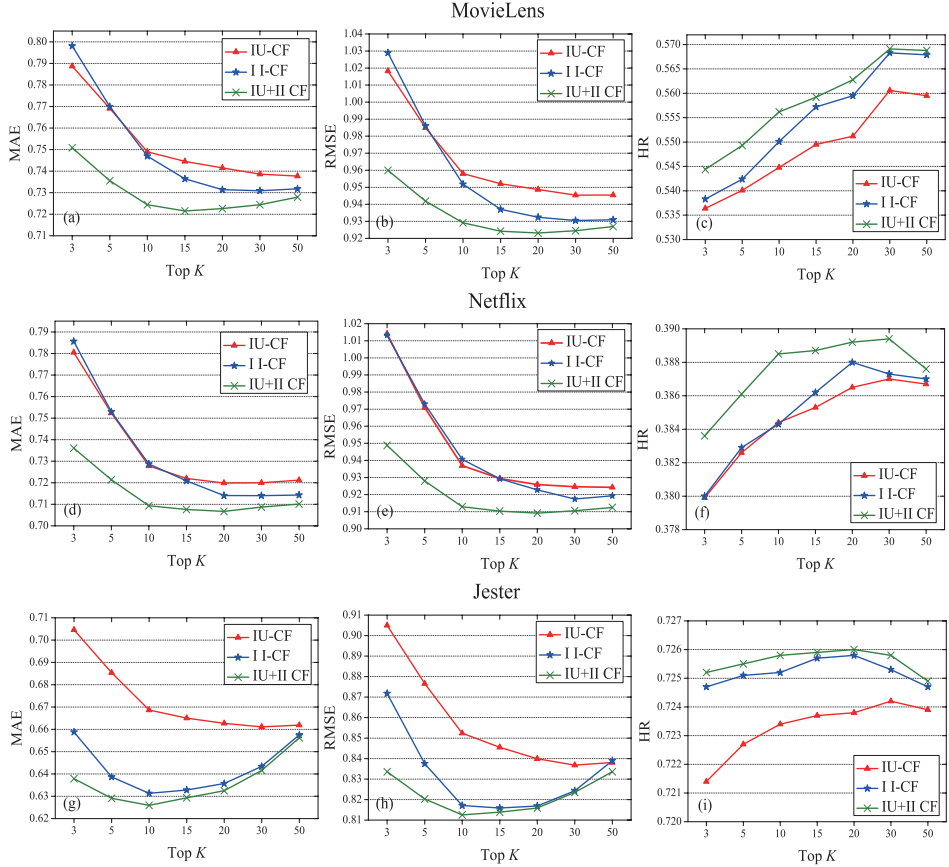}}
\caption{(Color Online) Comparison of results achieved by IU-CF,
II-CF and hybrid CF (IU+II CF) algorithms algorithms. Note that
hybrid CF algorithm has the smallest errors and highest hits.}
\label{Fig8}
\end{figure}

Figure \ref{Fig8} illustrates the comparison of IU-CF, II-CF and HCF
in the metrics MAE, RMSE, and HR respectively at the increasing
sizes of the neighborhood from 3 to 20 when the optimal parameter
$\alpha$ is 0.5. As shown in the Fig.\ref{Fig8}, for Movielens,
Netflix and Jester data sets, HCF obtains the remarkably lowest
levels of MAE and RMSE in the three methods when $K$ is quite small,
as well as highest HR values. Summing up the above three metric
results, the conclusion can reasonably be drawn that HCF which
integrates recommendations by implicit trust-based user and item
similarity network can further improve the performance of
recommendation in some degree than pure IU-F and II-CF. More
importantly, HCF could efficiently solve cold start problem.

%\frac{r_{ui}-\overline{r}_u}{R_u^{\max}-R_u^{\min}}-\frac{r_{vi}-\overline{r}_v}{R_v^{\max}-R_v^{\min}}
\section{Conclusion and discussions}
Information is explored dramatically in the social network era.
According to the structural properties of web connections, search
engineering could help us to dig out the most relevant web page
according to the keywords. However, search engineering couldn't help
users find the fresh information or products related to their
interests and habits, and couldn't analyze their personation,
either. Based on the user-item bipartite network, recommender system
is a promising tool to dig out the valuable information for the
users. However, the existing user or item correlation definition
didn't take into account the users' rating habits and statistical
properties in detail. Traditional CF algorithm suffers the
cold-start problem, and explicit trust-based recommender systems
require users to express explicit trust statements, which may be
time-consuming and expose privacy of users. Besides, the existing
implicit trust-based algorithms take few factors into consideration
to calculate the trust weight. Therefore, their recommendation
results are not sufficiently accurate. This work addresses these
problems by introducing implicit trust-based correlation network.
When computing implicit trust weight, we fully consider implicit
trust-based factors about users (e.g. average ratings, rating
ranges, and common experience) and items (e.g. internal attributes,
accept degrees and common popularity). The simulation results show
that the proposed implicit user-based, item-based and hybrid
algorithms solve cold start problem and provide accurate
recommendations.

Although our approaches presented in this article have shown
encouraging results, we also have several interesting tasks for
future work. First, we are going to focus on doing research on
transitive trust. In this paper, we have just paid attention to
computing the implicit trust weight, but have not studied trust
propagation. In real social network, trust can propagate from one
person to another. Due to trust propagation, perfect neighbors are
easy to be accessed and the cold start problem could also be
overcome in some degree. In the future, we are going to take
transitive trust into consideration in order to improve the
performance of implicit trust-based recommender system. Second, we
attempt to append robust mechanisms against the attacks by malicious
users to improve our proposed approaches. The reason is that some
e-commerce online recommender systems at present are often attacked
by negative canvassers. Therefore, it is worthwhile to emphasize the
robustness of an algorithm as an important aspect of practical
recommender systems. Finally, we plan to develop new evaluation
metrics to assess the performance of trust-based algorithms because
the current metrics seldom examine the robustness of recommender
systems.

% Do NOT remove this, even if you are not including acknowledgments
\section*{Acknowledgments}
We acknowledge the GroupLens Research Group for MovieLens data. This
work has been partly supported by the Natural Science Foundation of
China (Grant Nos. 10905052, 71171136, 71031002), the Fundamental
Research Funds for the Central Universities of China under Grant No.
DUT11RW422, and the Innovation Program of Shanghai Municipal
Education Commission (11ZZ135, 11YZ110). JGL is supported by the
Shanghai Leading Discipline Project (S30501), Shanghai Rising-Star
Program (11QA1404500) and Key Project of Chinese Ministry of
Education (211057).

%\section*{References}
% The bibtex filename
%\bibliography{template}

\newpage

%%%%%%%%%%%%%%%%%%%%%%%%%%%%%%

%\title{Supporting Information}
%\centerline{\LARGE Supporting Information}

%%%%%%%%%%%%%%%%%%%%%%%%%%%%%%%%%%%%%%%%%%%%%%%%%%%%%%%%%%%%%%%%

\end{document}